\documentclass[fleqn,12pt,twoside]{article}
\usepackage{espcrc1}

\usepackage{graphicx}
\usepackage[figuresright]{rotating}

\newcommand{\AmS}{{\protect\the\textfont2
  A\kern-.1667em\lower.5ex\hbox{M}\kern-.125emS}}

\title{Quark number susceptibilities from lattice QCD\thanks{Based on
work done \cite{qsus,fsus} with Sourendu Gupta and Pushan Majumdar.}
\vskip-2.0cm\hfill\small hep-ph/0110265, TIFR/TH/01-40\vskip1.5cm 
   }

\author{Rajiv V. Gavai\thanks{E-mail: gavai@tifr.res.in}\address{
                Department of Theoretical Physics, Tata
                Institute of Fundamental Research, \\
                Homi Bhabha Road, Mumbai 400 005, India}}

\begin{document}

% typeset front matter
\maketitle

\begin{abstract}
Results from our recent investigations of quark number susceptibilities
in both quenched and 2-flavour QCD are presented as a function of valence quark
mass and temperature.  A strong reduction ($\sim$40\%) is seen in the strange
quark susceptibility above $T_c$ in both cases.  A comparison of our isospin
susceptibility results with the corresponding weak coupling expansion reveals
once again the non-perturbative nature of the plasma up to $3T_c$.  Evidence
relating the susceptibility to another non-perturbative phenomena, 
pionic screening lengths, is presented.
\end{abstract}

\section{INTRODUCTION}

Investigations of quark number susceptibilities from first principles
can have direct experimental consequences since quark flavours such as electric
charge, strangeness  or baryon number can provide diagnostic tools for the
production of flavourless quark-gluon plasma in the central region of heavy ion
collisions.  As emphasized by B. M\"uller \cite{bm}, and Koch \cite{vk} in this
conference, fluctuations in conserved charges can act as probes of quark
deconfinement since they are different in the hadronic and plasma phases in
simple models.  Indeed, excess strangeness production has been suggested as a
signal of quark-gluon plasma almost two decades ago \cite{jrbm}.  Lattice QCD
can provide a very reliable and robust estimate for these quantities in {\em
both} the phases since in thermal equilibrium they are related to corresponding
susceptibilities by the fluctuation-dissipation theorem:
\begin{equation}
\langle \delta Q^2 \rangle \propto {T \over V} {{\partial^2 \log Z} \over 
\partial \mu_Q^2 } = \chi_Q(T, \mu_Q=0) ~.
\label{flds}
\end{equation}
Here $\mu_Q$ is the chemical potential for a conserved charge $Q$, and $Z$ is
the partition function of strongly interacting matter in volume $V$ at
temperature $T$.  Although Lattice QCD is unable to handle finite chemical
potential satisfactorily at present, and cannot thus yield any reliable
estimates of a number density, the susceptibility above, i.e., the first
derivative of the number density at zero chemical potential, can be obtained
reasonably well using the conventional simulation techniques.

Quark number susceptibilities also constitute an independent set of observables
to probe whether quark-gluon plasma is weakly coupled in the temperature regime
accessible to the current and future planned heavy ion experiments (say, 1 $
\le  T/T_c \le $ 10).  A lot of the phenomenological analysis of the heavy ion
collisions data is usually carried out assuming a weakly interacting plasma
although many lattice QCD results suggest otherwise.  It has been suggested
\cite{resum} that resummations of the finite temperature perturbation theory
may provide a bridge between phenomenology and the lattice QCD by explaining
the lattice results starting from a few $T_c$.  As we will see below, quark
number susceptibilities can act as a cross-check of the various resummation
schemes.  Earlier work on susceptibilities \cite{rly} did not attempt to
address this issue and were mostly restricted to temperatures very close to
$T_c$.  Furthermore, the quark mass was chosen there to vary with temperature
linearly.  We improve upon them by holding quark mass fixed in physical units
($m/T_c$ = constant). We also cover a larger range of temperature up to 3
$T_c$, and the accepted range of strange quark mass, in our simulations.

\section{FORMALISM}

After integrating the quarks out, the partition function $Z$ for QCD at finite
temperature and density is given by 
\begin{equation}
  Z = \int{\cal D}U {\rm e}^{-S_g}
            \det M(m_u,\mu_u)\det M(m_d,\mu_d)\det M(m_s,\mu_s).
\label{zqcd}
\end{equation}
Here $\{U_\nu(x)\}$, $\nu$ = 0--3, denote the gauge variables and 
$S_g$ is the gluon action, taken to be the standard Wilson action in 
our simulations.  Since we employ staggered fermions, the Dirac matrices,
$M$, are of dimensions 3$N_s^3 N_t$, with $N_s(N_t)$ denoting the number
of lattice sites in spatial(temporal) direction. $m_f$ and $\mu_f$ are quark
mass and chemical potential (both in lattice units) for flavour $f$, denoting
up(u), down(d), and strange(s) above.  The chemical potential needs to be
introduced on lattice as a function $g(\mu)$ and $g(-\mu)$ multiplying the gauge
variables in the positive and negative time directions respectively, such that
\cite{chem} i) $g(\mu) \cdot g(-\mu)$ = 1 and ii) the correct continuum limit 
is ensured.  While many such functions $g$ can be constructed, $\exp(\mu)$
being a popular choice, the results for susceptibilities at $\mu=0$ can easily
be shown to be independent of the choice of $g$ even for finite lattice spacing
$a$.  From the $Z$ in eq. (\ref{zqcd}), the quark number densities and the
corresponding susceptibilities are defined as
\begin{equation}
   n_f \equiv \frac{T}{V}\frac{\partial\ln Z}{\partial\mu_f} \qquad
%      = \frac{T}{V}
%            \left\langle{\rm tr} M_f^{-1} M_f'\right\rangle, \qquad
   \chi_{ff'} \equiv \frac{\partial n_f}{\partial \mu_{f'}}
             = \frac{T}{V}
      \left[\frac1Z\frac{\partial^2Z}{\partial\mu_f\partial\mu_{f'}}
          -\frac1Z\frac{\partial Z}{\partial\mu_f}\,
           \frac1Z\frac{\partial Z}{\partial\mu_{f'}}\right]~,
\label{incomplete}
\end{equation}
To lighten the notation, we shall put
only one subscript on the diagonal parts of $\chi$.
 
In order to obtain information for quark-gluon plasma in the central region,
we evaluate the susceptibilities at the point $\mu_f=0$ for all $f$.  In this
case, each $n_f$ vanishes, a fact that we utilize as a check on our numerical
evaluation.  Moreover, the product of the single derivative terms in eq.\
(\ref{incomplete}) vanishes, since each is proportional to a number density.
We set $m_u=m_d<m_s$.  Noting that staggered quarks have four flavours by
default, $N_f=4$, and defining $\mu_3 = \mu_u - \mu_d$, one finds from eq.
(\ref{incomplete}) that the isotriplet and strangeness susceptibilities are
given by 
\begin{equation}
\chi_3 = {T \over 2V} {\cal O}_1(m_u), \qquad
\chi_s = {T \over 4V} [{\cal O}_1(m_s) + {1 \over 4} {\cal O}_2(m_s)]~,~
\label{susc}
\end{equation}
where ${\cal O}_1 = \langle {\rm Tr} (M''M^{-1} - M'M^{-1}M'M^{-1}) \rangle$,
${\cal O}_2 = \langle ({\rm Tr} M'M^{-1})^2 \rangle$,  
$M'=\partial M/\partial\mu$ and $M''=\partial^2 M/\partial\mu^2$.  
The angular brackets denote averaging with respect to the $Z$ in 
eq. ({\ref{zqcd}).  One can similarly define 
baryon number and charge susceptibilities. We refer the reader for more details
on them to Ref. \cite{fsus}. 

In the discussion above, quark mass appears as an argument of ${\cal O}$ and
implicitly in the Boltzmann factor of $Z$.  Let us denote it by $m_{val}$ and
$m_{sea}$ respectively.  While the two should ideally be equal, we evaluated
the expressions above in steps of improving approximations (and increasing
computer costs) by first setting $m_{sea} = \infty $ for all flavours (quenched
approximation \cite {qsus}) and then simulating two light dynamical flavours,
by setting $m_{sea}/T_c = 0.1$ (2-flavour QCD \cite{fsus}).   In each case we
varied $m_{val}$ over a wide range to cover both light u,d quarks as well as
the heavier strange quark.  Details of our simulations as well as the technical
information on how the thermal expectation values of ${\cal O}$ were 
evaluated are in Refs. \cite{qsus,fsus}.

\section{RESULTS}

\begin{figure}[htb]
\vspace{-0.5cm}
\begin{center}
\includegraphics*[width=28pc]{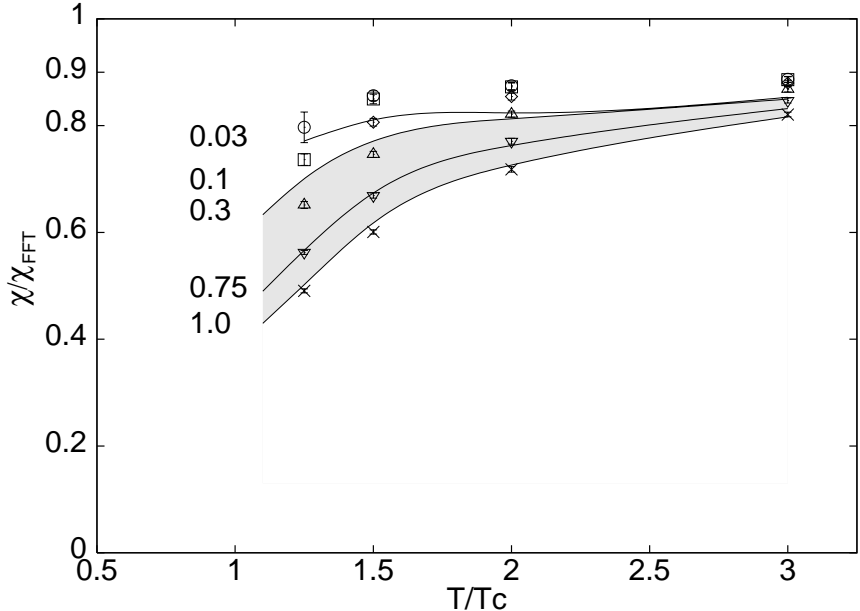}
\vspace{-0.5cm}
\caption{$\chi/\chi_{FFT}$ as a function of temperature for various
valence quark masses.}
\vspace{-0.5cm}
\label{fg.sus}
\end{center}
\end{figure}

Based on our tests \cite{qsus} of volume dependence, made by varying $N_s$ from
8 to 16, we chose $N_s =12$.  Choosing $N_t=4$, and making simulations at the
known $\beta_c(N'_t)$ for $N'_t = 6, 8, 12$, where $\beta_c$ is the gauge
coupling at which chiral (deconfinement) transition/cross-over takes place for
lattices with temporal extent $N'_t$,  we obtained results at $T/T_c$ =1.5, 2
and 3 in each case.  The sea quark mass in our dynamical simulations
corresponds to 14-17 MeV from the existing estimates \cite{tc} of $T_c$ for
2-flavour QCD.  Fig.  \ref{fg.sus} shows our results, normalized to the free
field values on the same size lattice, $\chi_{FFT}$, in the quenched
approximation (interpolated continuous lines) and for 2 light flavours (data
points) for a range of $m_{val}/T_c$ indicated on the left.  Due to our choice
of normalization, and the fact \cite{fsus} that the contribution of
${\cal O}_2$ to $\chi_s$ turns out to be negligibly small for $T > T_c$,
Fig. \ref{fg.sus} shows both $\chi_3$ and $\chi_s$.

Although $T_c$ differs in the quenched and 2-flavour QCD by a factor of
1.6-1.7, the respective susceptibilities in Fig. \ref{fg.sus} change by at most
5-10\% for each $m_{val}$.   This suggests that making further the strange
quark dynamical may not change the results significantly.  Using a wide range
for strange quark mass of 75 to 170 MeV, the strangeness susceptibility can be
read off from the shaded region.  It shows a strong temperature dependence and
is smaller by about 40\% compared to its ideal gas value near $T_c$. This has
implications for phenomenology of particle abundances. Finally, the ratio
$\chi/\chi_{FFT}$ is seen to be 0.88 (0.85) for 2-flavour (quenched) QCD at the
smallest $m_{val}$ and highest temperature we studied, with a mild 
temperature variation in its vicinity.

\begin{figure}[htb]
\vspace{-0.5cm}
\begin{center}
\includegraphics*[angle=270,width=25pc]{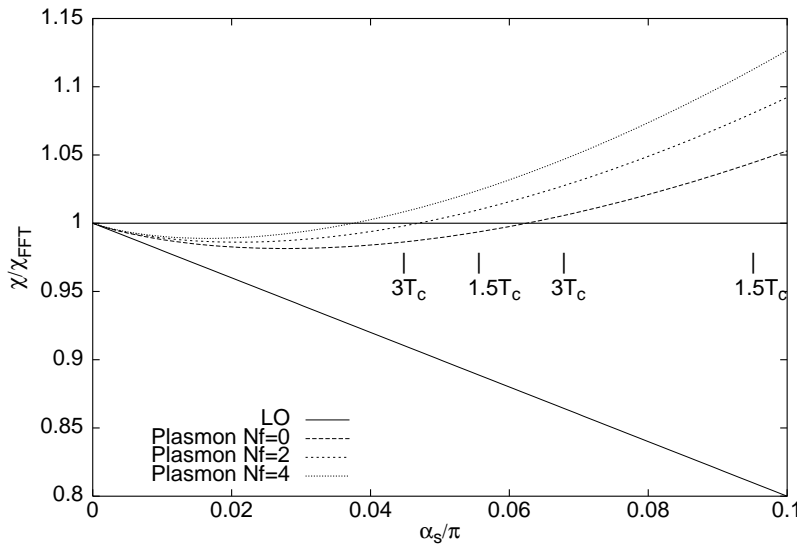}
\vspace{-0.5cm}
\caption{$\chi/\chi_{FFT}$ as a function of $\alpha_s/\pi$ for $N_f$ dynamical
massless quarks.}
\vspace{-0.5cm}
\label{fg.pert}
\end{center}
\end{figure}

Weak coupling expansion \cite{jk} yields $\chi/\chi_{FFT} = 1 - 2{\alpha_s
\over \pi} [1 - 4 \sqrt{{\alpha_s \over \pi}(1 + {N_f \over 6})}]$.  Fig.
\ref{fg.pert} shows the predictions for various $N_f$ along with the leading
order $N_f$-independent prediction.  Using a scale $2\pi T$ for the running
coupling and $T_c/\Lambda_{\overline{MS}} = 0.49 (1.15)$ for the $N_f = 2(0)$
theory \cite{tc}, the values $T/T_c$ =1.5 and 3 are marked on the figure as the
second (first) set.  As one can read off from the figure, in {\it both} cases
the ratio decreases with temperature in the range up to 3$T_c$ whereas our
results in Fig.  \ref{fg.sus} display an increase.  Furthermore, the
perturbative results lie significantly above in each case, being in the
range 1.027--1.08 for 2-flavour QCD and 0.986--0.994 for quenched QCD. This 
calls for clever resummations of perturbation theory or 
non-perturbative physics.

%\begin{minipage}[t]{77mm}
\begin{figure}[htb]
\vspace{-0.5cm}
\begin{center}
\includegraphics*[width=28pc]{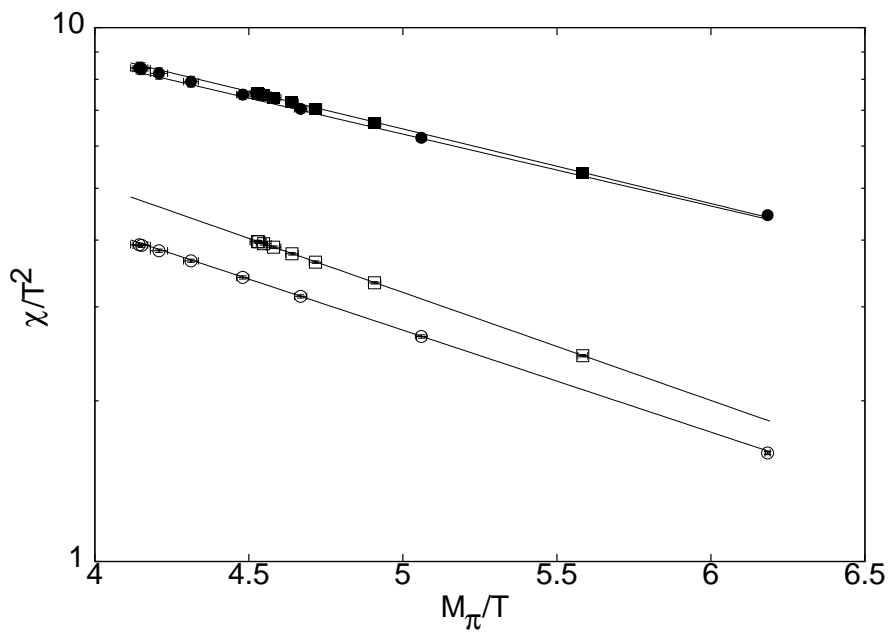}
\vspace{-0.6cm}
\caption{$4\chi_3/T^2$ (open symbols) and $\chi_\pi/10T^2$ (filled symbols) 
as a function of $M_\pi/T$ at $2T_c$ (circles) and $3T_c$ (boxes).}
\vspace{-0.7cm}
\label{fg.pi}
%\end{minipage}
\end{center}
\end{figure}
One known indicator of non-perturbative physics in the plasma phase is the
screening length in the channel with quantum numbers of pion.  While it 
exhibits chiral symmetry restoration above $T_c$ by being degenerate
with the corresponding scalar screening length, its value is much smaller than 
the free field value unlike that for other screening lengths.  Fig. \ref{fg.pi}
shows $\chi_3$ and $\chi_\pi$ (defined as a sum of the pion correlator over
the entire lattice) as a function of $M_\pi/T$.  It suggests the 
non-perturbative physics in the two cases to be closely related, if 
not identical.

\section{ACKNOWLEDGMENTS}
It is a pleasure to thank my collaborators Sourendu Gupta and Pushan Majumdar.
I am grateful to the Alexander von Humboldt Foundation for its generous
financial support which made my participation possible. It is a delight to
acknowledge the warm hospitality of the Physics Department of the University of
Bielefeld, especially that from Profs. Frithjof Karsch and Helmut Satz.


\begin{thebibliography}{9}
\bibitem{qsus} R. V. Gavai and S. Gupta, Phys. Rev. D 64 (2001) 074506.       
\bibitem{fsus} Rajiv V. Gavai, Sourendu Gupta and Pushan Majumdar,
{\tt hep-lat/0110032}.
\bibitem{bm} B. M\"uller, in these proceedings; M. Asakawa, U. W. Heinz,
and B. M\"uller, Phys. Rev. Lett. 85 (2000) 2072.
\bibitem{vk} V. Koch, in these proceedings; S. Jeon and V. Koch,
Phys. Rev. Lett. 85 (2000) 2076.
\bibitem{jrbm}J. Rafelski and B. M\"uller, Phys. Rev. Lett. 46 (1982) 1066;
erratum-ibid 56 (1986) 2334.
\bibitem{resum} 
   J. P. Blaizot {\sl et al.\/}, Phys. Rev. D 63 (2001) 065003;
   J. O. Andersen {\sl et al.\/}, Phys. Rev. D 63 (2001) 105008;
   K. Kajantie {\sl et al.\/}, Phys. Rev. Lett. 86 (2001) 10.        
\bibitem{rly} 
 S. Gottlieb {\sl et al.\/}, Phys. Rev. Lett. 59 (1987) 1513; 
 R. V. Gavai {\sl et al.\/}, Phys. Rev. D 40 (1989) 2743;
 S. Gottlieb {\sl et al.\/}, Phys. Rev. D 55 (1997) 6852.
\bibitem{chem} 
 R. V. Gavai, Phys. Rev. D 32 (1985) 519.
\bibitem{tc} 
   A. Ali Khan {\sl et al.\/}, Phys. Rev. D 63 (2001) 034502;
   F. Karsch {\sl et al.\/}, Nucl. Phys. B 605 (2001) 579;            
   S. Gupta, Phys. Rev. D 64 (2001) 034507.
\bibitem{jk} 
   J. I. Kapusta, {\sl ``Finite-temperature Field Theory''\/}, 1989,
      Cambridge University Press, Cambridge, UK, pp 132-133.     
\end{thebibliography}
\end{document}